\documentclass[10pt]{article}
\usepackage{spconf,amsmath,graphicx}

\usepackage{amssymb,comment,amsthm}

\usepackage{stfloats}
\usepackage{caption}
\usepackage{subcaption}
\usepackage{cite}
\usepackage{url}
\usepackage{algorithm}
\usepackage{algorithmic}

\usepackage{pgf}
\usepackage{pgfplots}
\usepackage{tikz}
\usepackage{subcaption}
\usepackage{eqparbox}
\usepackage{makecell}
\newcommand{\numAPs}{L}
\newcommand{\numAntennasPerAP}{N}
\newcommand{\numUEs}{K}

\newcommand{\taup}{\tau_{\rm p}}
\newcommand{\tauc}{\tau_{\rm c}}

\newcommand{\pilot}[1]{\boldsymbol{\phi}_{#1}}
\newcommand{\pilotMat}{\boldsymbol{\Phi}}

\newcommand{\chan}[1]{\mathbf{h}_{#1}}

\newcommand{\intchan}[1]{\mathbf{g}_{#1}}
\newcommand{\intchanEst}[1]{\widehat{\mathbf{g}}_{#1}}
\newcommand{\intsig}[1]{\mathbf{s}_{#1}}
\newcommand{\intsigProj}{\bar{\mathbf{s}}}
\newcommand{\intsigEst}[1]{\widehat{\bar{\mathbf{s}}}_{#1}}

\newcommand{\chanMat}[1]{\mathbf{H}_{#1}}
\newcommand{\chanMatEst}[1]{\widehat{\mathbf{H}}_{#1}}

\newcommand{\norm}[1]{\left\Vert#1\right\Vert}

\newcommand{\complexm}[2]{\mathbb{C}^{#1\times #2}}

\newcommand{\CN}[2]{\mathcal{CN}{(#1,#2)}}

\title{Distributed Signal Processing for Out-of-System Interference Suppression in Cell-Free Massive MIMO}
\name{Zakir Hussain Shaik and Erik G. Larsson \thanks{This work was partially funded by the REINDEER project of the European Union‘s Horizon 2020 research and innovation program under grant agreement No. 101013425. This work was also partially supported by ELLIIT and KAW.}}
\address{Department of Electrical Engineering (ISY), Linköping University, 581 83 Linköping, Sweden}
\begin{document}\ninept
%
\maketitle
\begin{abstract}
Cell-free massive multiple-input-multiple-output (CF-mMIMO) is a  next-generation wireless access technology that offers superior coverage and spectral efficiency compared to conventional MIMO. 
With many future applications in unlicensed spectrum bands, networks will likely experience and may even be limited by out-of-system (OoS) interference. 
The OoS interference differs from the in-system interference from other serving users in that for OoS interference, the associated pilot signals are unknown or non-existent, which makes estimation of the OoS interferer channel difficult. 

In this paper, we propose a novel sequential algorithm for the suppression of  OoS interference for uplink CF-mMIMO with a stripe (daisy-chain) topology.  
The proposed method has comparable performance to that of a fully centralized interference rejection combining algorithm but has substantially less fronthaul load requirements.
\end{abstract}
\section{Introduction}
\label{sec:intro}
Cell-free massive multiple-input-multiple-output (CF-mMIMO) is envisaged to be one of the main next-generation physical-layer technology \cite{emil2020competitive}. In CF-mMIMO, many distributed access points (APs) simultaneously serve each user equipment (UE) in the network. An AP is a transceiver circuitry comprising antenna elements and signal processing units required for local processing.

5G offers, and 6G is envisioned to offer, configurations for operation in  unlicensed spectrum \cite{parkvall20205g}. 
Although this opens up for a wide range of applications, it also brings new signal-processing challenges \cite{zhang2017survey,ahmad20205g}. 
One challenging aspect is that if the network operates in  unlicensed spectrum or shares the spectrum with another system, then there will be inherent out-of-system (OoS) interference. 
This OoS interference differs from in-system interference that arises from other UEs in that the OoS interference signals are completely unknown: they may or may not contain pilots, and even if they contain pilots, those pilots are unknown.
These challenges get manifold with CF-mMIMO because the OoS interference affects different APs through different channels, making coherent interference rejection difficult.
Throughout this paper, the focus is on CF-mMIMO.

In the literature,  interference rejection is commonly achieved by pre-whitening the received signal with an estimated interference-plus-noise covariance matrix and treating the resulting signal as if it were interference-free.
This approach goes back at least to the 1990's \cite{karlsson1996interference,winters1993signal} and some later works can be found in \cite{craig2002system,hsieh2011low,sabbagh2021cell,huang2019fronthaul}. 
Most of the cited papers, either process all signals centrally or process signals in a decentralized manner at each AP locally without inter-AP cooperation.
The local processing approach will have poorer performance because it does not exploit a vital property of OoS interference: that the unknown interfering signal is the same at all  APs, although it experiences different channels.  
On the other hand, processing all the signals centrally at the central processing unit (CPU) entails a substantial fronthaul load for non-centralized CF-mMIMO topologies like  radio stripes \cite{interdonato2019ubiquitous}, for example.
The radio stripe is a sequential (daisy-chain) CF-mMIMO topology and share the same cables for fronthaul and power supply. The recent works on stripe topology have focused on developing algorithms where there is no OoS interference. For instance, the Kalman inspired algorithm \cite{zakir2021mmse} achieves optimal performance in the sense of mean-square-error (MSE). 

The main focus of this paper is to develop interference rejection algorithms for a sequentially distributed CF-mMIMO network, which to the best of our knowledge, existing literature did not address. The specific contributions of this paper are: we propose a sequential processing using Gramian matrices to achieve the equivalent performance as centralized processing with less fronthaul load, and we propose a novel algorithm that has comparable performance as the Gramian method but with substantially lower fronthaul load. In the payload transmission phase, we treat the OoS interference source as an additional fictitious user that we detect and discard.

\section{System model and channel Estimation}\label{SystemModel}
We consider a CF-mMIMO network with stripe topology comprising $\numAPs$ APs, each equipped with $\numAntennasPerAP$ antennas serving $\numUEs$ single-antenna UEs. We also assume that there is single unknown OoS interfering source equipped with a single antenna. We model the channels between the users (including the OoS interfering source) to the APs as block fading channels and denote by $\chan{kl}$ the fading channel between UE $k$ and AP $l$. We further assume that the channels are unknown a priori to both APs and UEs. We consider that system is operating in time-division duplex (TDD) mode for channel estimation and payload transmission. The APs estimate the downlink channels from the pilot signals transmitted by UEs in uplink.

For the purpose of channel estimation, we assume that there are $\numUEs$ mutually orthogonal $\tau_p \geq \numUEs$ length pilot signals and let the pilot assigned to UE $k$, for $ k=1,\ldots, K$, be denoted by $\pilot{k}$ with $\Vert \pilot{k} \Vert^2=1$. With the assumptions made, we receive the following signal during the pilot phase at AP $l$ over $\taup$ channel uses:
\begin{equation}
	\begin{aligned}
		\mathbf{Y}_l &= \sqrt{p\tau_p} \chanMat{l}\pilotMat^H + \intchan{l}\intsig{}^H + \mathbf{N}_l,
	\end{aligned}
\end{equation}
where $p$ is the transmit power of each UE, $\chanMat{l} = [\chan{1l},\ldots,\chan{Kl}]$ is the channel matrix between the UEs and AP $l$, $\pilotMat = [\pilot{t_1},\ldots,\pilot{t_{K}}]$ is the pilot matrix, $\intchan{l}$ is the channel between the OoS interfering source and AP $l$, $\intsig{}$ is the transmit signal of the OoS source over $\taup$ channel uses and $\mathbf{N}_l$ is the receiver noise at AP $l$ where we model each entry of the matrix as i.i.d. circular Gaussian distributed with variance $\sigma^2$.

For the channel estimation, we employ least squares (LS) and accordingly we compute the channel estimate $\chanMatEst{l}$ at AP $l$ as follows
\begin{equation}\label{chanEstAPl}
	\begin{aligned}
		\chanMatEst{l} = \frac{1}{\sqrt{p\tau_p}}	\mathbf{Y}_l\pilotMat.
	\end{aligned}
\end{equation}

\section{Interference Channel Estimation} \label{sec:channEst}
To estimate the channel of the OoS source, we first preprocess the received signal at each AP to remove or minimize from it the channel component of the serving users, $\{\chanMat{l}, l =1,\ldots,L\}$. We shall call such a preprocessed received signal in this paper ``residual signal". We compute the residual signal at AP $l$, denoted by $\mathbf{Z}_l\in \complexm{N}{\taup}$, by observing that the LS channel estimate in \eqref{chanEstAPl} can be expressed as
\begin{equation}
	\begin{aligned}
		\chanMatEst{l}  =\chanMat{l} + \frac{1}{\sqrt{p\tau_p}}\left(\intchan{l}\intsig{}^H + \mathbf{N}_l\right)\pilotMat
	\end{aligned}
\end{equation}
and then subtracting the known signal component from the received signal as follows:
\begin{equation}\label{resSign}
	\begin{aligned}
		\mathbf{Z}_l &= \mathbf{Y}_l - \sqrt{p\tau_p} \chanMatEst{l}\pilotMat^H\\
		& = \left(\intchan{l}\intsig{}^H + \mathbf{N}_l\right)\left(\mathbf{I}-\pilotMat\pilotMat^H\right)\\
		& = \left(\intchan{l}\intsig{}^H + \mathbf{N}_l\right)\mathbf{P},
	\end{aligned}
\end{equation}
where $\mathbf{P} = \mathbf{I}-\pilotMat\pilotMat^H$ is the projection onto the orthogonal compliment of the pilot matrix $\pilotMat$. To make any meaningful estimate of $\intsig{}$ from the residual matrices, we require that $\taup > K$, otherwise, the residual matrices at all the APs will be zero. It is worth noting that  due to the non-invertibility of the projection matrix in \eqref{resSign}, the OoS interfering signal $\mathbf{s}$ cannot be estimated completely. However, we can instead estimate its component spanned by $\mathbf{P}$, i.e., $\mathbf{P}\intsig{}$. Against this background, we decompose the projection matrix as 
\begin{equation}\label{Pdecomp}
	\mathbf{P} = \mathbf{\Psi} \mathbf{\Psi}^H,
\end{equation}
where $\mathbf{\Psi}$ is a tall matrix that satisfies the orthogonality property $\mathbf{\Psi}^H\mathbf{\Psi} = \mathbf{I}$. This decomposition is the economy-size-singular value decomposition (SVD) of $\mathbf{P}$. Note that $\mathbf{P}$ is of dimension $\taup \times \taup$ and has rank $\taup -K$. Therefore, $\mathbf{\Psi}$ is of dimension $\taup \times (\taup-K)$. We will denote in our further discussions 
\begin{equation} \label{projIntsignal}
	\intsigProj = \mathbf{\Psi}^H\mathbf{s}
\end{equation}
as the projected signal component that we can estimate without ambiguity. To obtain the estimate $\intsigEst{}$ of $\intsigProj$, first, we despread the residual signal by projecting it on $\mathbf{\Psi}$ as follows
\begin{equation}\label{procResidual}
	\begin{aligned}
		\mathbf{Z}_l\mathbf{\Psi} &= \left(\intchan{l}\intsig{}^H + \mathbf{N}_l\right)\mathbf{\Psi}\\
		&=\intchan{l} \intsigProj^H + \mathbf{N}_l^{'},
	\end{aligned}
\end{equation}
where $\mathbf{N}_l^{'}=\mathbf{N}_l \mathbf{\Psi}$ is an $\numAntennasPerAP \times (\taup-K)$ noise matrix, whose entries are i.i.d. because $\mathbf{\Psi}$ is unitary matrix \cite{redbook}. There is no loss of information with the projection in \eqref{procResidual} because we can recover the original residual signal in \eqref{resSign} by multiplying \eqref{procResidual} by $\mathbf{\Psi}^H$. Having obtained the processed signal at each AP, $\{\mathbf{Z}_l\mathbf{\Psi}, l =1,\ldots,L\}$, we will present in the following subsections different methods to compute the OoS source channel and signal estimates. These methods trade off between fronthaul load and performance. In discussions to follow, we define fronthaul load quantitatively as the number of real symbols transmitted in each link connecting two APs.

\subsection{Centralized processing as a baseline for comparison} \label{centImpl}
In a centralized processing, we collect the processed residual signals from all the APs at the CPU and solve the following LS problem to obtain the estimate of $\intsigEst{}$  

\begin{equation} \label{optForm1}
	\underset{\intchan{},\bar{\mathbf{s}}}{\rm{minimize}} \quad  \norm{\mathbf{Z}\mathbf{\Psi} - \intchan{}\intsigProj^H}_F,
\end{equation}
where $\mathbf{Z} = [\mathbf{Z}_1^T,\ldots,\mathbf{Z}_L^T]^T$, $\intchan{}= [\intchan{1}^T,\ldots,\intchan{L}^T]^T$ and $\norm{\cdot}_F$ is the Frobenius norm of the argument. 

We can obtain a global solution by taking the best rank-1 approximation of $\mathbf{Z}\mathbf{\Psi}$ using SVD, i.e., estimates of $\intchan{}$ and $\bar{\mathbf{s}}$ are left and right singular vectors of $\mathbf{Z}\mathbf{\Psi}$ scaled by largest singular value. Centralized processing will perform better than any other scheme because the CPU has access to the processed residual signals from all the APs. However,
to sequentially accumulate and forward all the processed residuals, $\mathbf{Z}_l\mathbf{\Psi}$, will incur a significant fronthaul load, especially in the link connecting AP $L$ and the CPU. The fronthaul load in the link between AP $l$ and AP $l+1$ is $2N(\taup-K) l$. Any processing scheme whose fronthaul load increases proportionally with the number of APs is undesirable in practice for stripe topologies. Although central processing is undesirable, it forms a baseline in terms of performance for any scheme that improves on fronthaul load.

In the following subsections, we will present three different processing methods for sequential networks. The first two methods form a baseline in terms of minimum fronthaul requirement and performance, respectively. The third method is the proposed novel method that trades off between fronthaul load and performance.

\subsection{Method 1: Local processing at each AP}\label{localProcMethod1}
We can achieve zero fronthaul load with local processing at each AP without cooperating with other APs. AP $l$ makes use of the locally processed residual signal in \eqref{procResidual} to computes the best local LS estimates of $\intchan{l}$ and $\intsigProj$ by solving the following problem
\begin{equation} \label{localEst1}
	\underset{\intchan{l},\bar{\mathbf{s}}}{\rm{minimize}} \quad  \norm{\mathbf{Z}_l\mathbf{\Psi} - \intchan{l}\intsigProj^H}_F.
\end{equation}
One way to obtain a solution is to compute the best rank-1 approximation of $\mathbf{Z}_l\mathbf{\Psi}$ through SVD similar to centralized processing discussed in Section \ref{centImpl}. However, these estimates in general will be suboptimal because the APs do not take into consideration that $\intsig{}$ is the same at all APs. Moreover, in local processing APs does not exploit the network's topology to cooperate with other APs.

\subsection{Method 2: Sequential accumulation of Gramians}
One of the main problems with centralized processing is that fronthaul load increases with the number of APs. We can circumvent this issue without loss in performance by sequentially accumulating Gramians. To understand this method, observe from Section \ref{centImpl} that a global estimate of $\intsigEst{}$ of $\intsig{}$ is the right singular vector of $\mathbf{Z}\mathbf{\Psi}$ or the dominant eigenvector of $\mathbf{\Psi}^H\mathbf{Z}^H\mathbf{Z}\mathbf{\Psi}$. The later expression $\mathbf{\Psi}^H\mathbf{Z}^H\mathbf{Z}\mathbf{\Psi}$ is the Gramian of the matrix $\mathbf{Z}\mathbf{\Psi}$. Also, note the following alternative computation of the Gramian
\begin{equation}\label{Gram}
	\mathbf{\Psi}^H\mathbf{Z}^H\mathbf{Z}\mathbf{\Psi} = \mathbf{\Psi}^H\left(\sum_{l=1}^{L}\mathbf{Z}_l^H\mathbf{Z}_l\right)\mathbf{\Psi}.
\end{equation}
With this discussion in the background, we can compute the Gramian by sequentially accumulating local Gramians through sequential fronthaul. Specifically, AP $l$ receives $\sum_{i=1}^{l-1}\mathbf{Z}_i^H\mathbf{Z}_i$ from AP $l-1$, and adds its local Gramian $\mathbf{Z}_l^H\mathbf{Z}_l$, and forwards to the consecutive AP in the stripe. Finally, the CPU computes the estimate $\intsigEst{}$, which is sent back to all the APs. After receiving the estimate $\intsigEst{}$, AP $l$ estimates its corresponding OoS interfering channel as 
\begin{equation}\label{LocChanEst}
	\intchanEst{l} = \norm{\intsigEst{}}^{-2}\mathbf{Z}_l\mathbf{\Psi}\intsigEst{},\ l =1,\ldots,L.
\end{equation}
Although the Gramian-based method achieves the same performance as a centralized processing, it suffers from one main drawback that each AP has to forward a complete Gramian matrix sequentially, amounting to an increase in the fronthaul load. The fronthaul requirement between each link is $\taup^2$. Note that from \eqref{Gram}, each AP can directly add and forward the Gramian obtained from processed residual in \eqref{procResidual}, then the fronthaul load is $(\taup - K)^2$.

\subsection{Method 3: Sequential phase rotation and averaging}
Having seen two sequential methods with different favorable features: one with least fronthaul load but with suboptimal performance and other with optimal performance but with large fronthaul load, we now present a novel approach that has an attractive trade off between fronthaul load and performance. The first step involves AP $l$ computing a local estimate of the OOS interfering signal similar to Method 1 in Section \ref{localProcMethod1}. Since the interfering signal is the same at all APs, in principle, one can obtain an estimate by simply averaging all the local estimates $\intsigEst{l}, l=1,\ldots, L$. However, the dominant singular vector of a matrix is not unique, and it is ambiguous up to an arbitrary phase rotation. Hence, AP $l$ can only estimate $\intsigProj$ up to an unknown phase rotation, and direct averaging may not work in general. We tackle this phase ambiguity by having AP $l$ compute the average of the $\intsigProj$-estimate received from AP $l-1$ and adding its local estimate with a phase-rotation that is optimal in the sense of LS. Intuitively, we phase-align the two estimates. Mathematically, this process is done as follows
\begin{equation}
	\intsigEst{l} = 0.5\left(\intsigEst{(l-1)} + \widehat{\mathbf{s}}^{\rm o}_le^{j\alpha_l}\right),
\end{equation}
where $\intsigEst{l}$ is the final estimate at AP $l$, $\widehat{\mathbf{s}}_l^{\rm o}$ is the right singular vector of $\mathbf{Z}_l\mathbf{\Psi}$ at AP $l$, the rotation angle is computed as 
\begin{equation}
\alpha_l = -\arg\left(\intsigEst{(l-1)}^H\widehat{\mathbf{s}}^{\rm o}_l\right).
\end{equation}
At the first AP, we can either arbitrary initialize or initialize with $\intsigEst{0}=\mathbf{0}$.

\begin{figure}[!htb]
	\centering
	\includegraphics[ height=2cm, keepaspectratio]{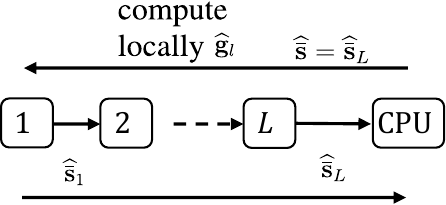}
	\caption{Demonstrating sequential phase rotation and averaging}
	\label{fig:spra}
\end{figure}
The fronthaul load in each link between the APs is $2(\taup - K)$ real symbols. The fronthaul cost of this scheme is minimal compared to centralized processing and Method 2. Finally, each AP computes the local interfering channel estimates as in \eqref{LocChanEst}. A flowchart of the method is shown in Fig. \ref{fig:spra}.

\begin{table*}
	\normalsize
	\centering
	\begin{tabular}{ |p{5cm}|p{3.5cm}|p{4.5cm}| }
		\hline
		\multicolumn{3}{|c|}{\textbf{Fronthaul load in each coherence block}} \\
		\hline
		\textbf{Method/Algorithm}&\textbf{Channel estimation phase} &\textbf{Payload phase with LS estimator}\\
		\hline
		\makecell[l]{Centralized Processing \\(between AP $L$ and the CPU)}&$2N(\taup - K) L$   &$2NL + 2N(K+1)L$\\
		\hline
		Method 1: Local processing&   $0$  & $2(K+1) + (K+1)^2$\\
		\hline
		Method 2: Gramian based processing &$(\taup-K)^2$ & $2(K+1) + (K+1)^2$\\
		\hline
		Method 3: Phase rotate and average   &$2(\taup - K)$ & $2(K+1) + (K+1)^2$\\
		\hline
	\end{tabular}
	\caption{Summary of fronthaul signaling load for various algorithms} \label{tab_fronthaul}
	\vspace{-4mm}
\end{table*}

\section{Payload Transmission}\label{sec:payload}
During the payload phase, data can be detected either using central processing (maximum-ratio, zero-forcing, or minimum-mean-square-error (MMSE) estimator) or using any decentralized processing known in the art, e.g., \cite{zakir2021mmse}. In either case, the OoS interference signal is simply treated as an additional unknown fictitious user, which is eventually discarded.

At AP $l$ we receive the following signal in a symbol period
\begin{equation}
	\begin{aligned}
		\mathbf{y}_l &= \mathbf{H}_l\mathbf{x} + \mathbf{g}_ls + \mathbf{n}_l\\
		&=\begin{bmatrix}
			{\mathbf{H}}_l & {\mathbf{g}}_l
		\end{bmatrix} \begin{bmatrix}
			\mathbf{x}\\ s
		\end{bmatrix} + \mathbf{n}_l,\\
	\end{aligned}
\end{equation}
where $\mathbf{x}$ is the payload transmitted by UEs, $s$ is the OoS sample (scalar) and $\mathbf{n}_l$ is the receiver noise at AP $l$ with Gaussian distribution. We consider centralized LS as the estimation technique to demonstrate the performance of the methods presented in Section \ref{sec:channEst}.

\subsection{Centralized LS processing}\label{sec:centLS}
With the centralized processing, we jointly estimate the desired signal $\mathbf{x}$ and the OoS interference signal $s$ using LS (zero-forcing) as follows:
\begin{equation}\label{centLS}
	\begin{bmatrix}
		\widehat{\mathbf{x}}\\ \widehat{s}
	\end{bmatrix} =  \begin{bmatrix}
		\widehat{\mathbf{H}} & \widehat{\mathbf{g}}
	\end{bmatrix}^{\dagger}\mathbf{y},\vspace{-2mm}
\end{equation}
where $\widehat{\mathbf{x}}$ and $\widehat{s}$ are the estimates of $\mathbf{x}$ and $s$, respectively, $\mathbf{y}= [\mathbf{y}_1^T,\ldots,\mathbf{y}_L^T]^T$ is the augmented received signal from all APs and $(\cdot)^\dagger$ is the pseudoinverse of the argument. The fronthaul load is $2NL + 2N(K+1)L$ i.e., the final AP has to forward all the signals and channel estimates (including that of OoS interference source) it accumulates from all the other APs sequentially to the CPU.

\subsection{Sequential LS}
For all the sequential methods presented in Section \ref{sec:channEst}, we use sequential LS as the estimation technique \cite{kay1993fundamentals}. The LS estimate of the desired signal and the OoS interfering signal at AP $l$ is given by

\begin{equation} \label{SeqLS}
	\begin{bmatrix}
		\widehat{\mathbf{x}}_l\\ \widehat{s}_l
	\end{bmatrix} =  \begin{bmatrix}
		\widehat{\mathbf{x}}_{l-1}\\ \widehat{s}_{l-1}
	\end{bmatrix} + \mathbf{T}_l(\mathbf{y}_l - \widehat{\mathbf{H}}_l \widehat{\mathbf{x}}_{l-1} - \widehat{\mathbf{g}}_l s_{l-1}), l =1,\ldots, L,
\end{equation}
where \vspace{-2mm}
{\footnotesize
\begin{align}
		\mathbf{T}_l &= \mathbf{Q}_{l-1}\begin{bmatrix}
			\widehat{\mathbf{H}}_l & \widehat{\mathbf{g}}_l
		\end{bmatrix}^H\left(\sigma^2\mathbf{I} + \begin{bmatrix}
			\widehat{\mathbf{H}}_l & \widehat{\mathbf{g}}_l
		\end{bmatrix} \mathbf{Q}_{l-1}\begin{bmatrix}
			\widehat{\mathbf{H}}_l & \widehat{\mathbf{g}}_l
		\end{bmatrix}^H\right)^{-1},\\
	\mathbf{Q}_l &= \left(\mathbf{I} - \mathbf{T}_l \begin{bmatrix}
		\widehat{\mathbf{H}}_l & \widehat{\mathbf{g}}_l
	\end{bmatrix}\right)\mathbf{Q}_{l-1},
\end{align}
}
and initial values $\begin{bmatrix}
	\widehat{\mathbf{x}}_0\\ \widehat{s}_0
\end{bmatrix} = \begin{bmatrix}
	{\mathbf{0}}\\ {0}
\end{bmatrix}$ and $\mathbf{Q}_{0} = 	\alpha \mathbf{I}$ with $\alpha$ being some large positive constant, this is to avoid biasing the estimator towards initial estimate $\intsigEst{0}$ \cite{kay1993fundamentals}. Also, matrices $\{\mathbf{Q}_l\}$ are positive semi-definite.

We summarize the fronthaul load for all the methods in Table \ref{tab_fronthaul}. We observe that among sequential methods, Method 1 has the least fronthaul load and Method 2 the largest.

\section{Numerical Results and Discussions}\label{numericalResults}
In this section, we evaluate the performance of the proposed algorithm through numerical results, and the metric of performance considered is bit-error-rate (BER). One can also use mean-square error (MSE), however the relative performance does not change. The channel model is the Rayleigh fading, i.e., the channel between UE $k$ and AP $l$ is
\begin{equation}
	\chan{kl} \sim \CN{\mathbf{0}}{\beta_{kl}\mathbf{I}},
\end{equation}
where $\beta_{kl}$ is the large-scale fading coefficient. However, importantly, the methodology presented in the channel estimation and payload phase is applicable to any channel model. For large-scale fading, we consider the 3GPP Urban Microcell propagation model \cite[Table~B.1.2.1-1]{LTE2010b} with $2$ GHz carrier frequency and according to it, the coefficients are given as follows
\begin{equation}
	\beta_{kl} = -30.5 - 36.7\textrm{log}_{10}\left(\frac{{d}_{kl}}{1\textrm{m}}\right),
\end{equation} 
where $d_{kl}$ is the distance between AP $l$ and UE $k$ (this includes a vertical height difference of $5$ m between the APs and the UEs). We consider a uniform linear array at each AP with half-wavelength antenna spacing. The number of APs considered is $L = 4$, the number of antennas per AP is $N=4$, the number of UEs $K=5$, and there is one OoS interference source. We consider a simulation setup of a $500\times 500$ square grid with APs being equally spaced on the border. We deploy all the users, including the OoS source, randomly within the concentric square grid of the setup with $10$ m gap from the border. We use QPSK modulation during the payload phase. We consider $\tauc = 200$ and $\taup = 50$ channel uses. Typically, for low mobility scenarios, $\tauc$ is much larger than $200$ and, therefore, will have minimal effect on the net spectral efficiency of the network.

\begin{figure}[H] 
	\centering
	\begin{tikzpicture}
		\begin{semilogyaxis}[
			width=1\linewidth,
			height=0.8\linewidth,
			ymin = 1e-3,			
			xmin = -10,
			xmax = 0,
			legend pos = south west,
			legend cell align={left},
			legend columns=1, 
			xlabel = {Uplink power (normalized) [dB]},
			ylabel = {BER}
			]
			\addplot[thick, blue, solid,mark=o,mark size=2pt] table [x index = 0, y index = 1, col sep=comma] {Figures/Plot1Data.dat};
			\addplot[thick, red, solid,mark=square,mark size=2pt] table [x index = 0, y index = 2, col sep=comma] {Figures/Plot1Data.dat};
			\addplot[thick, blue, solid,mark=+,mark size=2pt] table [x index = 0, y index = 3, col sep=comma] {Figures/Plot1Data.dat};
			\addplot[thick, black, dashed,mark=asterisk,mark size=2pt] table [x index = 0, y index = 4, col sep=comma] {Figures/Plot1Data.dat};	
			\addplot[thick, red, solid,mark=diamond,mark size=2pt] table [x index = 0, y index = 5, col sep=comma] {Figures/Plot1Data.dat};
			\addlegendentry{No Int. Suppression};
			\addlegendentry{Seq.  Local Processing};
			\addlegendentry{Seq.  Phase Rotation (proposed)};
			\addlegendentry{Seq.  Gramian Based (proposed)};
			\addlegendentry{Cent. Genie};
		\end{semilogyaxis}
	\end{tikzpicture}
	\caption{Experimental results indicating the benefits of the proposed algorithm}
	\label{plot1}\vspace{-2mm}
\end{figure}
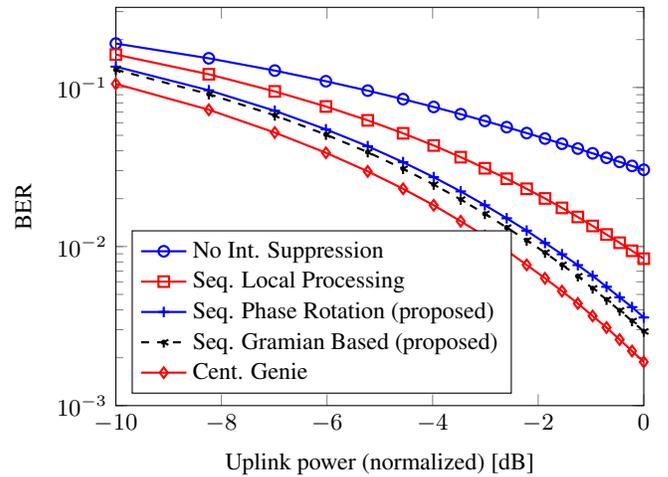

We use a nearest-point detector on the processed signal i.e., on \eqref{centLS} for centralized processing and on \eqref{SeqLS} for sequential methods. The Fig. \ref{plot1} shows the BER performance of the different methods. The results are obtained by averaging over many setups of random user (including OoS source) locations. The interfering source signal is drawn from a Gaussian distribution with $-3$ dB normalized power. The description of different curves in the figure is as follows:

\begin{itemize}
	\item \textbf{No interference suppression}: This is the baseline without any attempts to suppress the OoS interference. This could be implemented either with centralized or decentralized processing, e.g., using the Kalman filter approach\cite{zakir2021mmse}.
	\item \textbf{Sequential local processing (Method 1)}: Each AP forms local estimates of the channels to the OoS interfering source. This method does not exploit the fact that all APs see the same OoS interference signal.
	\item \textbf{Sequential Gramian based (Method 2)}: This is the proposed distributed OoS interference suppression method using the accumulation of the Gramians method for the OoS interference channel estimation. This method is performance-wise equivalent to centralized processing discussed in Sections \ref{centImpl} and \ref{sec:centLS}.
	\item \textbf{Sequential phase rotation (Method 3)}: This is the proposed distributed OoS interference suppression method using the sequential-phase-rotation method for the OoS interference channel estimation. Note the slight performance loss compared to the Gramian-based approach, while savings in fronthaul are substantial.
	\item \textbf{Centralized genie detector}: This is the baseline (genie) case where centralized processing is done with perfect knowledge of all channels of the UEs and the OoS interfering source, i.e., $\{\mathbf{H}_l, \mathbf{g}_l\}, l=1,\ldots,L$.
\end{itemize}

From Fig. \ref{plot1}, it is evident that if the network does not suppress the OoS interference and ignores it, then there is a significant loss in performance. The Gramian-based approach has the best performance among the sequential cases presented, and that is because its performance is equivalent to that of centralized processing (not shown in the plot to avoid redundancy). Local processing has the lowest fronthaul load requirement among all the sequential processing algorithms. However, the performance is poorer. From this numerical result, the proposed phase rotation and average method has a strikingly good trade off among sequential methods, i.e., it has comparable performance to the Gramian-based approach with significantly less fronthaul load.
\vspace{-2mm}
\section{Conclusion}
This paper proposed algorithms to estimate and suppress the OoS interference in CF-MIMO with sequential topology. The proposed novel phase-rotate-and-average algorithm exploits the fact that interference is the same at all APs. Moreover, the numerical simulation demonstrates that the phase-rotate-and-average algorithm has comparable performance to a centralized implementation with substantially less fronthaul load, making it a suitable candidate for stripe topologies. In future work, we intend to extend the work to multiple/high-rank OoS interference sources.

\bibliographystyle{IEEEtran}
\bibliography{ZakRef.bib}

\end{document}